\documentclass[preprint2]{aastex63}

\newcommand{\kms}{{km s$^{-1}$}}

\newcommand{\myemail}{jskang@astro.snu.ac.kr}
\newcommand{\profemail}{mglee@astro.snu.ac.kr}

\usepackage{amsmath}
\hypersetup{linkcolor=red,citecolor=blue,filecolor=cyan,urlcolor=magenta}

\accepted{May 18, 2020}
\submitjournal{ApJ}

\shorttitle{M64-GC1 and RGB stars}
\shortauthors{Kang et al.}

\begin{document}

\title{
A New Metal-poor Globular Cluster and Resolved Stars in the Outer Disk of the Black Eye Galaxy M64: 
Implication for the Origin of the Type III Disk Break
}

\correspondingauthor{Myung Gyoon Lee}
\email{\myemail,\profemail}

\author[0000-0003-3734-1995]{Jisu Kang} 
\affiliation{Astronomy Program, Department of Physics and Astronomy, Seoul National University, 
1 Gwanak-ro, Gwanak-gu, Seoul 08826, Republic of Korea}

\author[0000-0003-1392-0845]{Yoo Jung Kim} 
\affiliation{Astronomy Program, Department of Physics and Astronomy, Seoul National University, 
1 Gwanak-ro, Gwanak-gu, Seoul 08826, Republic of Korea}

\author[0000-0003-2713-6744]{Myung Gyoon Lee} 
\affiliation{Astronomy Program, Department of Physics and Astronomy, Seoul National University, 
1 Gwanak-ro, Gwanak-gu, Seoul 08826, Republic of Korea}

\author[0000-0002-2502-0070]{In Sung Jang} 
\affiliation{Leibniz-Institut f{\"u}r Astrophysik Potsdam (AIP), An der Sternwarte 16, 
14482 Potsdam, Germany}

\begin{abstract}
{
M64 is a nearby spiral galaxy with a Type III anti-truncation component. 
To trace the origin of the Type III component,
we present {\it Hubble Space Telescope}/Advanced Camera for Surveys $F606W/F814W$ photometry of resolved stars in a field located in the outer disk ($2\farcm5 \lesssim r \lesssim 6\farcm5$) of M64.
At $r\approx 5\farcm5$ (7 kpc) to the east, 
we discover a new metal-poor globular cluster ($R_{\rm eff}=5.73\pm0.02$ pc and $M_V=-9.54\pm0.09$ mag), M64-GC1.
This is the first globular cluster found in M64. 
The color-magnitude diagram (CMD) of the resolved stars in M64-GC1 is well matched by 12 Gyr isochrones with [Fe/H] $=-1.5\pm0.2$, 
showing that this cluster belongs to a halo. 
The CMD of the resolved stars in the entire ACS field shows two distinguishable red giant branches (RGBs): a curved metal-rich RGB and a vertical metal-poor RGB.
The metal-rich RGB represents an old metal-rich ([Fe/H] $\approx -0.4$) disk population. 
In contrast, the CMD of the metal-poor RGB stars is very similar to the CMD of M64-GC1, 
showing that the metal-poor RGB represents a halo population.
The radial number density profile of the metal-rich RGB stars is described by an exponential disk law, 
while the profile of the metal-poor RGB stars is described by a de Vaucouleurs's law.
From these, we conclude that the origin of the Type III component in M64 is a halo which has a much lower metallicity than a disk or bulge population.
} 
\end{abstract}



\section{Introduction}\label{intro}

M64 (NGC 4826) is an early type spiral galaxy ((R)SA(rs)ab) 
with nicknames of Black Eye Galaxy, Evil Eye Galaxy, and Sleeping Beauty Galaxy 
due to its prominent dust lane around the bright bulge. 
The basic properties of M64 are listed in {\color{blue}\bf Table \ref{tab_m64}}.

\begin{deluxetable}{lcl}
\tablewidth{0pc} 
\tablecaption{Basic Parameters of M64\label{tab_m64}}
\tablehead{
\colhead{Parameter} 	& \colhead{Value} & \colhead{Ref.}}
\startdata
R.A.(J2000)         	& 12$^h$ 56$^m$ 43$^s$.64              & 1 \\
Decl.(J2000)        	& +21$\arcdeg$ 40$\arcmin$ 58$\farcs$7 & 1 \\
Type 		& (R)SA(rs)ab	                       & 2 \\
$A_B$, $A_V$, $A_I$  & 0.150, 0.113, 0.062   & 3 \\
$A_{F606W}$, $A_{F814W}$ & 0.102, 0.063      & 3 \\
$E(B-V)$ & 0.037                              & 3 \\
$(m-M)_0$	& $28.18 \pm 0.03({\rm ran}) \pm 0.08({\rm sys})$   & 4 \\
Distance                    & $4.33 \pm 0.18$ Mpc                           & 4 \\
Image scale 				& 21.0 pc arcsec$^{-1}$                         & 4 \\
                            & 1.26 kpc arcmin$^{-1}$                        & 4 \\
                            & 75.6 kpc degree$^{-1}$                         & 4 \\
$B^T$, $(B-V)^T$  	& $9.36\pm0.10$, $0.84\pm0.01$ & 2 \\
$B^T_0$, $(B-V)^T_0$ & $8.82\pm0.10$, $0.71\pm0.01$ & 2 \\
$M_B^T$, $M_V^T$ & $-19.36\pm0.13$, $-20.07\pm0.13$  & 2,4 \\
Position angle 			& 115 deg $(B)$, 110 deg $(K_s)$ & 2 \\
$D_{25}(B)$				& $600\farcs00 \times 324\farcs00$ & 2 \\
$D_{tot}(K_s)$			& $617\farcs60 \times 352\farcs03$ & 5 \\ 
$v_{helio}$ & $408\pm4$ km s$^{-1}$ & 2 \\
\enddata
\tablerefs{(1) NED (2) \citet{dev91} (3) \citet{sch11} (4) This study (5) \citet{jar03}}
\end{deluxetable}

M64 shows several interesting features. 
First, it is an isolated galaxy located in the low density environment.
It used to be known to have only one known companion dwarf galaxy, NGC 4789A ($M_B^T\approx -14$ mag) \citep{jac09}.
Although several dwarf galaxies including Coma P ($M_B^T\approx -10$ mag) were recently suggested to be another companions of M64 \citep{bru19,car20}, 
it can be still considered as an isolated galaxy.
Second, it hosts two counter-rotating gas rings: 
the inner gas ring ($r<1'$) follows the galaxy rotation, 
while the outer gas ring rotates in the opposite direction \citep{bra92}. 
The outer ring is considered to be a remnant of a recent merger with a gas-rich satellite 
\citep{bra92,bra94,rub94,wal94,rix95,wat16}.
Third, recent star formation is concentrated only in the inner disk region at $r<200''$, 
while the HI distribution is much more extended (out to $r\approx 600''$) \citep{wat16}.


Another interesting feature of M64 is that its disk is composed of three substructures.
From the recent deep $BV$ surface photometry, 
\citet{wat16} found that the surface brightness profiles of M64 can be divided into three parts:
1) an inner disk, 2) an outer disk, and 3) a Type III anti-truncation (upbending) component.
In the inner disk region at $r<200''$, spiral arms and evidence of recent star formation are seen.
We can notice dust lanes, young stars, strong UV emission, and a high density of HI gas from the multi-band images of M64 \citep[see][Figure 3]{wat16}.
The recent star formation in the inner disk is believed to be due to a merger with a gas-rich dwarf galaxy \citep{bra94,wat16}. 
In the outer disk region at $200''<r<400''$, no spiral arms or evidence of recent star formation are seen.
Its color is red, 
indicating that the disk stars may be old and metal-rich \citep{wat16}.
Outside of the outer disk at $400''<r<900''$, the Type III anti-truncation component is located.
It is an upbending structure following a break at $r\approx 400''$ in the surface brightness profiles of M64.
Its surface brightness is as low as $\mu_B>27$ mag arcsec$^{-2}$.





While the origins of the inner disk and the outer disk are quite clear,
the origin of the Type III break is still mysterious.
According to \citet{wat16}, the origin of the Type III component in M64 can be explained by three scenarios. 
First, it can be originated from disk stars that migrated from the inner region.
This possibility is supported by the $(B-V)$ color of the Type III component stars
which is as red as that of the outer disk stars.
However, the uncertainty of the color measurement for $r>400''$ is significant because of its low surface brightness.
Moreover, it is difficult for disk stars to migrate such a large distance (several disk scale lengths beyond the spiral features).
Second, the Type III component can be originated from star formation in extended gas 
induced by counter rotating kinematics of M64.
If so, the $(B-V)$ color of the Type III component stars may be much bluer (younger) than previously known.
Third, it can be originated from halo stars.
If so, the stars in this component must be old and metal-poor.


Photometry of resolved stars located in the low surface brightness region of a galaxy is an excellent tool to study the nature of diffuse galaxy light in terms of stellar populations 
and to trace the origin of the stellar populations.
In this study, we explore the origin of the Type III break 
using the photometry of resolved stars in an outer disk field of M64 
based on the {\it Hubble Space Telescope (HST)}/Advanced Camera for Surveys (ACS) images.
To date, little is known about the resolved stars or star clusters in M64.
We adopt the distance to M64, 4.33 Mpc as determined later in this study.
The foreground reddening toward M64 is as low as $E(B-V)=0.037$ 
($A_{F606W}=0.102$ and $A_{F814W}=0.063$) \citep{sch11}.

This paper is organized as follows.
Section 2 describes data used in this study and how we reduce and analyze the data. 
In Section 3, we present main results:
a discovery of a new globular cluster,
color-magnitude diagrams (CMDs) of the resolved field stars, 
estimation of the tip of the red giant branch (TRGB) distance,
and radial number density profiles of the resolved stars.
Implications of our results are discussed in Section 4. 
Finally, we summarize our results and give a conclusion.

\section{Data and Data Reduction}\label{data}


We used archival data of {\it HST}/ACS $F606W$/ $F814W$ images of M64 (PI: Tully, PID: 10905).
We combined individual \texttt{flc} (charge transfer efficiency corrected) frames of each filter using \texttt{DrizzlePac} \citep{gon12}.
The total exposure times are 922 s for $F606W$ and 1128 s for $F814W$.
The final image scale is $0\farcs05$ per pixel.
 
\begin{figure} 
\centering
\includegraphics[scale=0.48]{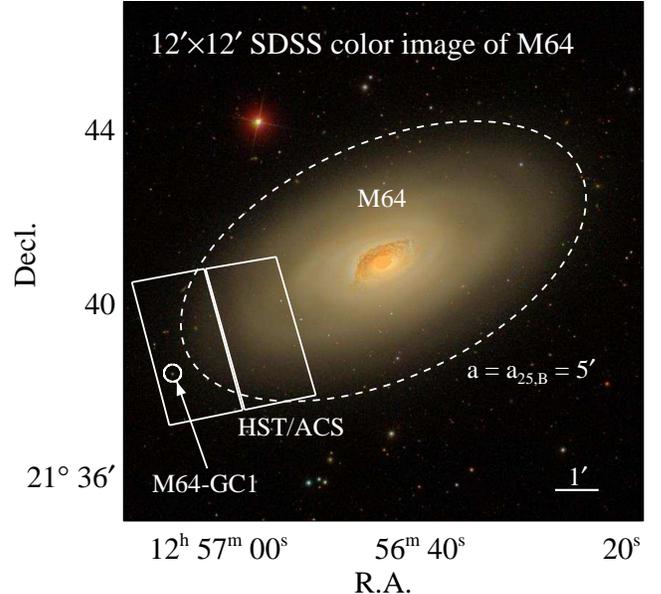} 
\caption{
The color map of the $12'\times 12'$ SDSS image of M64.
North is up and east to the left.
The image scale bar is shown on the bottom right.
The large ellipse represents $a_{25,B}=5\farcm0$ 
where $a_{25,B}$ is the projected galactocentric distance at which the $B$-band surface brightness is 25 mag arcsec$^{-2}$. 
The location of the {\it HST}/ACS field is marked in rectangles. 
The small circle on the {\it HST}/ACS field marks the new globular cluster (M64-GC1) found in this study (see {\color{blue}\bf Figure \ref{fig_fig3}}).
}
\label{fig_fig1}
\end{figure}

In {\color{blue}\bf Figure \ref{fig_fig1}}, the location of the {\it HST}/ACS field is marked on the SDSS color image of M64.
The ACS field is positioned at $r=2\farcm3$ to $6\farcm2$ ($a=2\farcm4$ to $6\farcm6$)
along the eastern major axis from the center of M64. 
Here, $r$ is the galactocentric distance and $a$ is the projected galactocentric distance.
Note that the galaxy light of the SDSS image is clearly seen in the western chip of the ACS, while it is barely seen in the eastern chip.

\begin{figure} 
\centering
\includegraphics[scale=0.48]{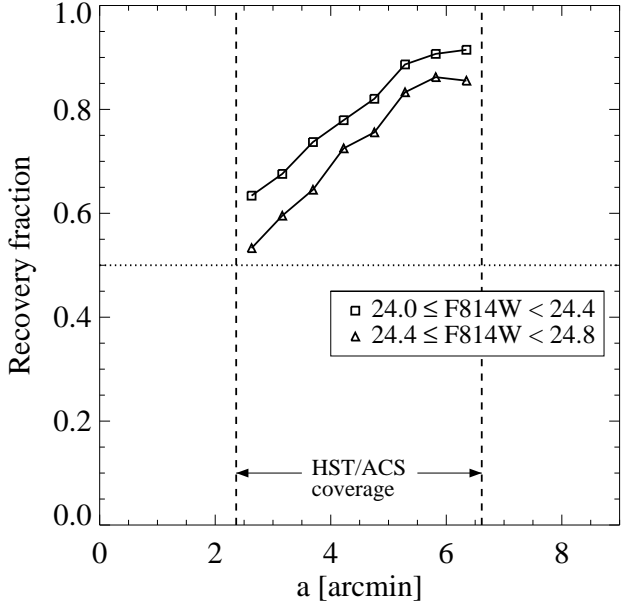}
\caption{
The recovery fraction of the bright stars along the radial bin 
according to the two different magnitude range ($24.0\le F814W<24.4$, $24.4\le F814W<24.8$).
This result was derived from the artificial star tests.
The vertical dashed lines mark the {\it HST/ACS} coverage and the horizontal dotted line marks the 50\% completeness. 
}
\label{fig_fig2}
\end{figure}


For the detection and photometry of the point sources in the images, 
we utilized \texttt{DOLPHOT} \citep{dol00}.
The drizzled $F814W$ image was used as a reference image, 
and the individual \texttt{flc} images were used as input images. 
We used \texttt{DOLPHOT} input parameters listed in Table A2 of \citet{mon16}
which were used for the photometry of GHOSTS (Galaxy Halos, Outer disks, Substructure, Thick disks, and Star clusters) fields.
We applied selection criteria for the stars similar to those used in the GHOSTS survey \citep{rad11}: 
--0.06 $<$ \texttt{SHARP}$_{F606W}$ $+$ \texttt{SHARP}$_{F814W}$ $<$ 1.30, 
\texttt{CROWD}$_{F606W}$ $+$ \texttt{CROWD}$_{F814W}$ $<$ 1.0, 
\texttt{S/N}$_{F606W,F814W}$ $>$ 5, 
\texttt{type}$_{F606W,F814W}$ $=$ 1, 
\texttt{flag}$_{F606W,F814W}$ $\le$ 2. 
We increased the maximum crowding parameter from 0.16 to 1.0 considering the crowding of the M64 field.
We use the Vega magnitude system following the \texttt{DOLPHOT} output parameters.
Aperture corrections were automatically applied by \texttt{DOLPHOT} routine (\texttt{Apcor} = 1). 
The systematic errors of aperture corrections are on average $\pm0.03$ mag \citep{mon16}.


Artificial star tests were also carried out using the \texttt{DOLPHOT} routine \texttt{fakelist} to generate the star lists.
We generated 1,000 artificial stars with a magnitude range of $23.5<F606W<27.5$ and a color range of $0.5<(F606W-F814W)<3.0$,
and re-ran \texttt{DOLPHOT} with FakeStar parameters 
(\texttt{FakeMatch} = 3, \texttt{FakePSF} = 2, \texttt{FakeStarPSF} = 1, 
\texttt{RandomFake} = 1, \texttt{FakePad} = 0).
We repeated this procedure 100 times to generate 100,000 artificial stars in total.
{\color{blue}\bf Figure \ref{fig_fig2}} shows 
recovery fraction of the bright stars along the radial bin 
according to the two different magnitude ranges ($24.0\le F814W<24.4$, $24.4\le F814W<24.8$).
The recovery fraction decreases with decreasing projected galactocentric distance $a$, 
but it is higher than 50\% even at the innermost region.
This result was used to correct the radial number density profile of the RGB stars in {\color{blue}\bf Section \ref{rdp}}.

\section{Results}

\subsection{Discovery of a New Globular Cluster}

\begin{figure} 
\centering
\includegraphics[scale=0.48]{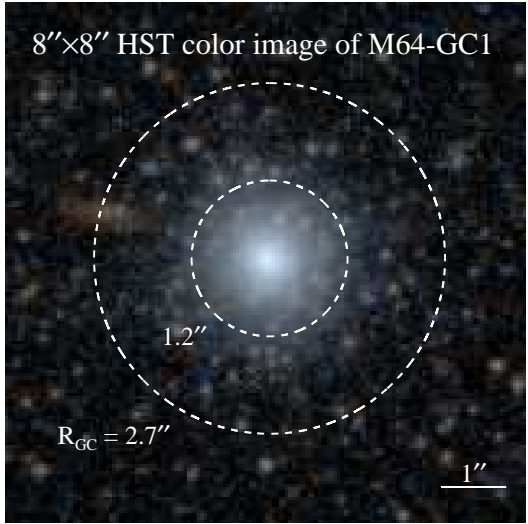} 
\caption{
The color map of the $8''\times 8''$ zoomed-in {\it HST} image of M64-GC1. 
North is up and east to the left.
The image scale bar is shown on the bottom right.
Note that the outer region of M64-GC1 is partially resolved into individual stars. 
They are mostly RGB stars of the cluster. 
Dashed lines represent the circles with radii of $R_{GC}=1\farcs2$ and $2\farcs7$. 
}
\label{fig_fig3}
\end{figure}

From the visual inspection of the ACS images, 
we 
discovered a new globular cluster, henceforth named M64-GC1, located at $r\approx 5\farcm5$ in the east. 
This is the first globular cluster found in M64, 
and it is the only star cluster with resolved stars we could find in the images. 
The position of M64-GC1 is marked by a circle in {\color{blue}\bf Figure \ref{fig_fig1}}. 
This cluster is seen as a bluish point source in the SDSS color image.
{\color{blue}\bf Figure \ref{fig_fig3}} is a pseudo color map of the zoomed-in {\it HST}/ACS image for M64-GC1.
This new cluster shows a very round shape and its outer region is partially resolved into individual stars, which verifies that it is indeed a star cluster.

\begin{figure} 
\centering
\includegraphics[scale=0.48]{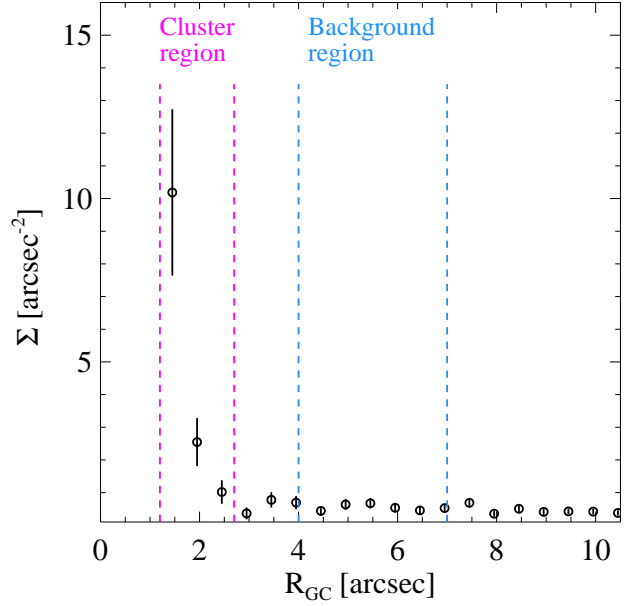} 
\caption{
The radial number density profile of the detected stars with $F814W<26.5$ mag around M64-GC1. 
No stars are detected in the central region at $R_{GC}<1\farcs2$.
Magenta lines at $R_{GC}=1\farcs2$ and $2\farcs7$ show the boundary for the cluster region, 
and pale blue lines at $R_{GC}=4\farcs0$ and $7\farcs0$ denote the background region adopted in this study.
}
\label{fig_fig4}
\end{figure}

We derived the radial number density profile of the resolved stars by counting the detected stars with $F814W<26.5$ mag in the star cluster region at $R_{GC}<10''$ from the cluster center, as displayed in {\color{blue}\bf Figure \ref{fig_fig4}}.
No stars are detected in the central region at $R_{GC}<1\farcs2$ due to severe crowding. 
However, the radial profile clearly shows an excess at $1\farcs2<R_{GC}<2\farcs7$ 
and a flat distribution in the outer region at $R_{GC}>4''$. 
This shows that most detected stars at $1\farcs2<R_{GC}<2\farcs7$ are the members of the cluster.

\begin{figure*} 
\centering
\includegraphics[scale=0.48]{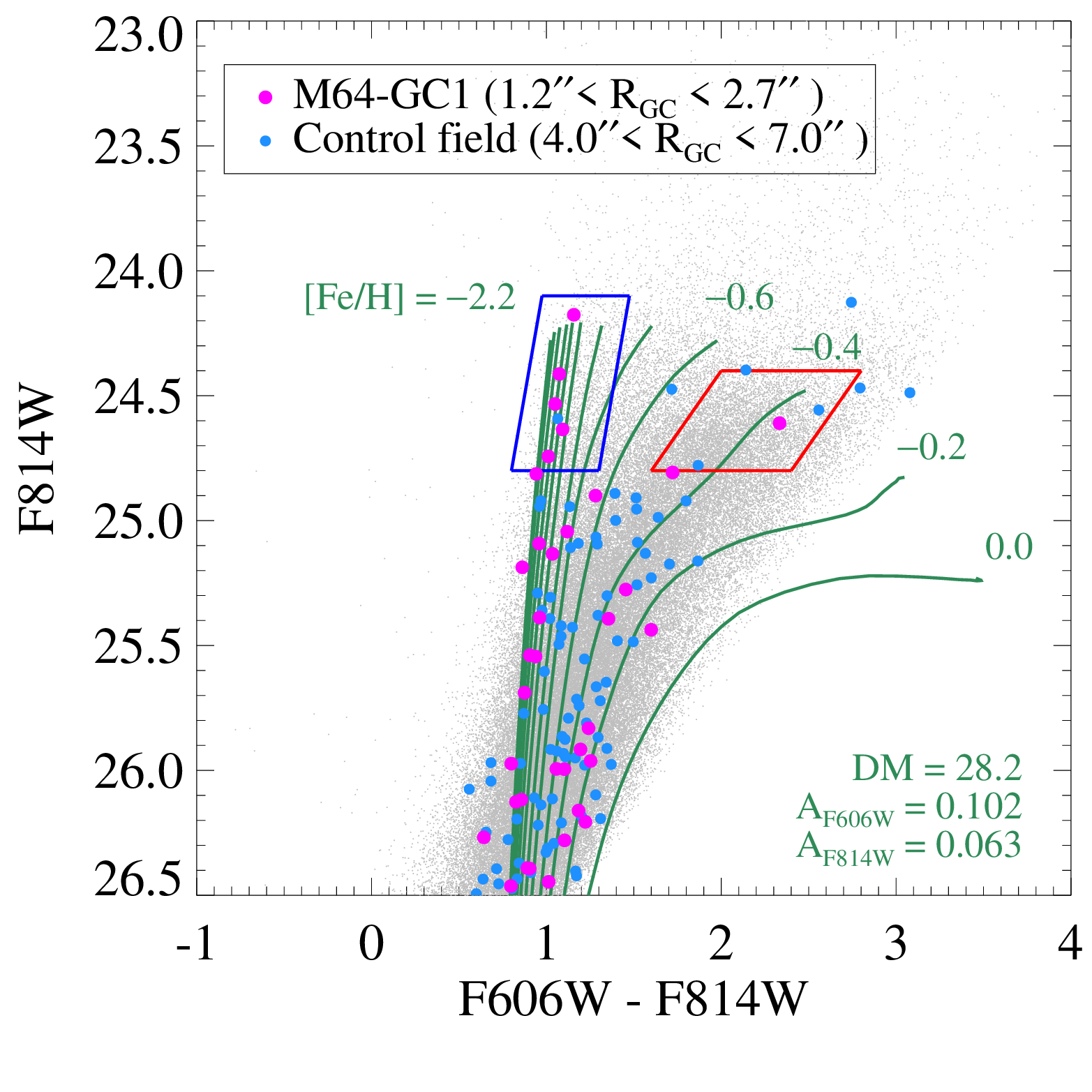}
\includegraphics[scale=0.48]{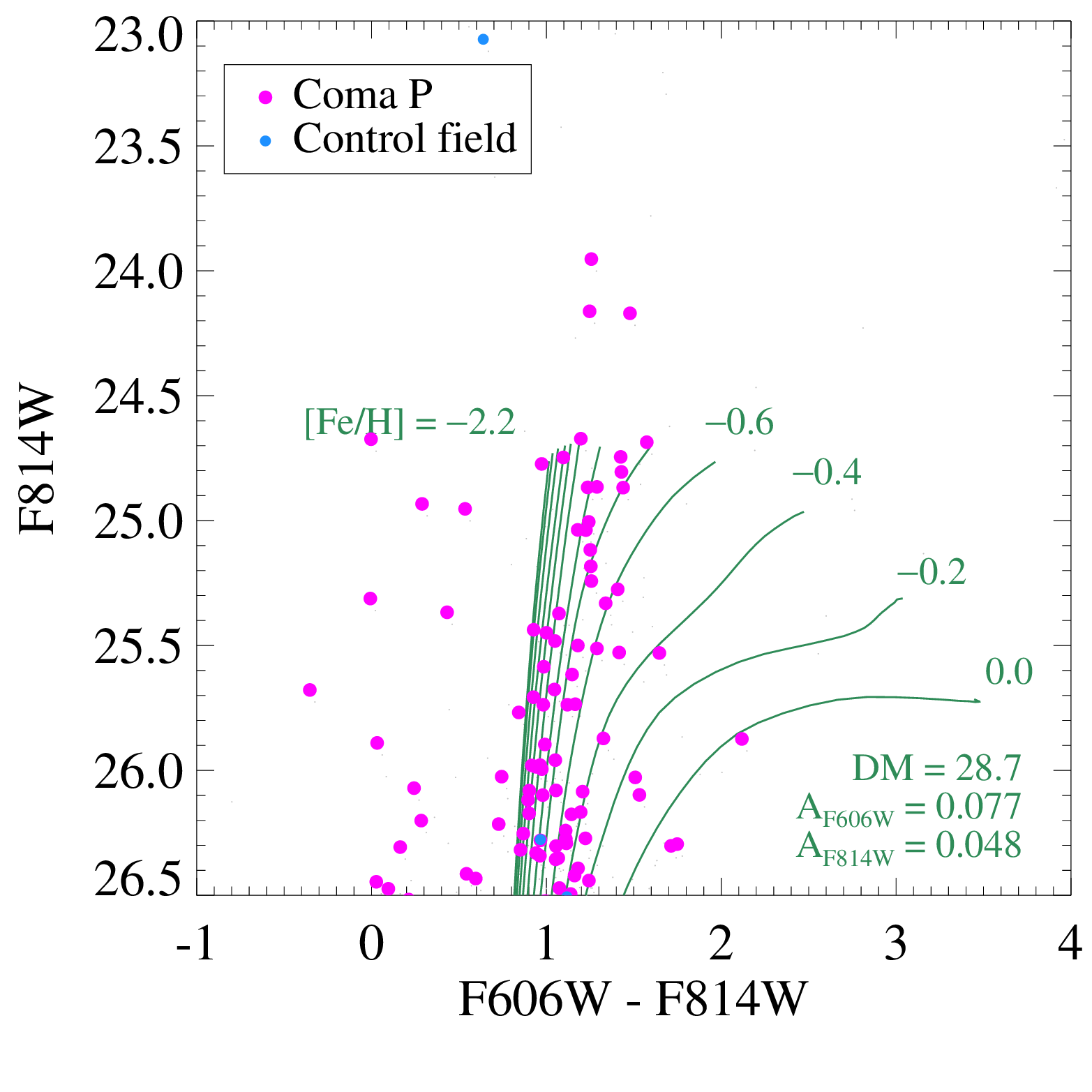}
\caption{
(Left) CMD of the resolved stars in M64-GC1. 
Magenta circles mark the stars at $1\farcs2<R_{GC}<2\farcs7$, pale blue circles mark the background stars at $4\farcs0<R_{GC}<7\farcs0$, and grey dots mark the stars in the entire ACS field.
Green lines denote isochrones for age = 12 Gyr and [Fe/H]$=-2.2$ to 0.0 with a step of 0.2 dex, 
shifted according to $(m-M)_0=28.2$, $A_{F606W}=0.102$, and $A_{F814W}=0.063$. 
Blue and red boxes denote the boundaries for selecting bright metal-poor RGB stars and metal-rich RGB stars later in this study.
(Right) CMD of the resolved stars in Coma P. 
Magenta circles for the Coma P stars, and pale blue circles for the background stars.
Green lines denote the same isochrones shifted according to $(m-M)_0=28.7$, $A_{F606W}=0.077$, and $A_{F814W}=0.048$. 
}
\label{fig_fig5}
\end{figure*}

\begin{figure*} 
\centering
\includegraphics[scale=0.48]{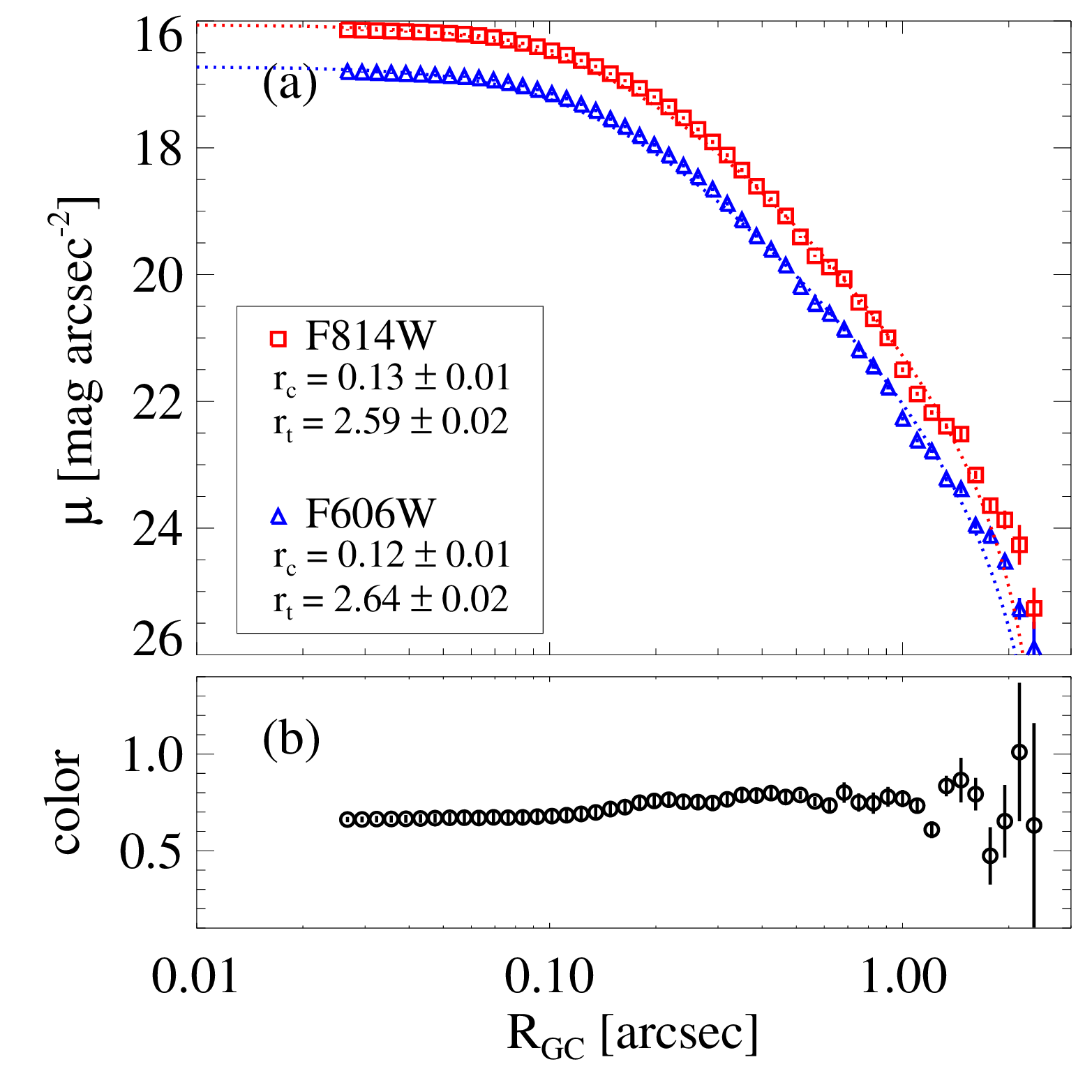} 
\includegraphics[scale=0.48]{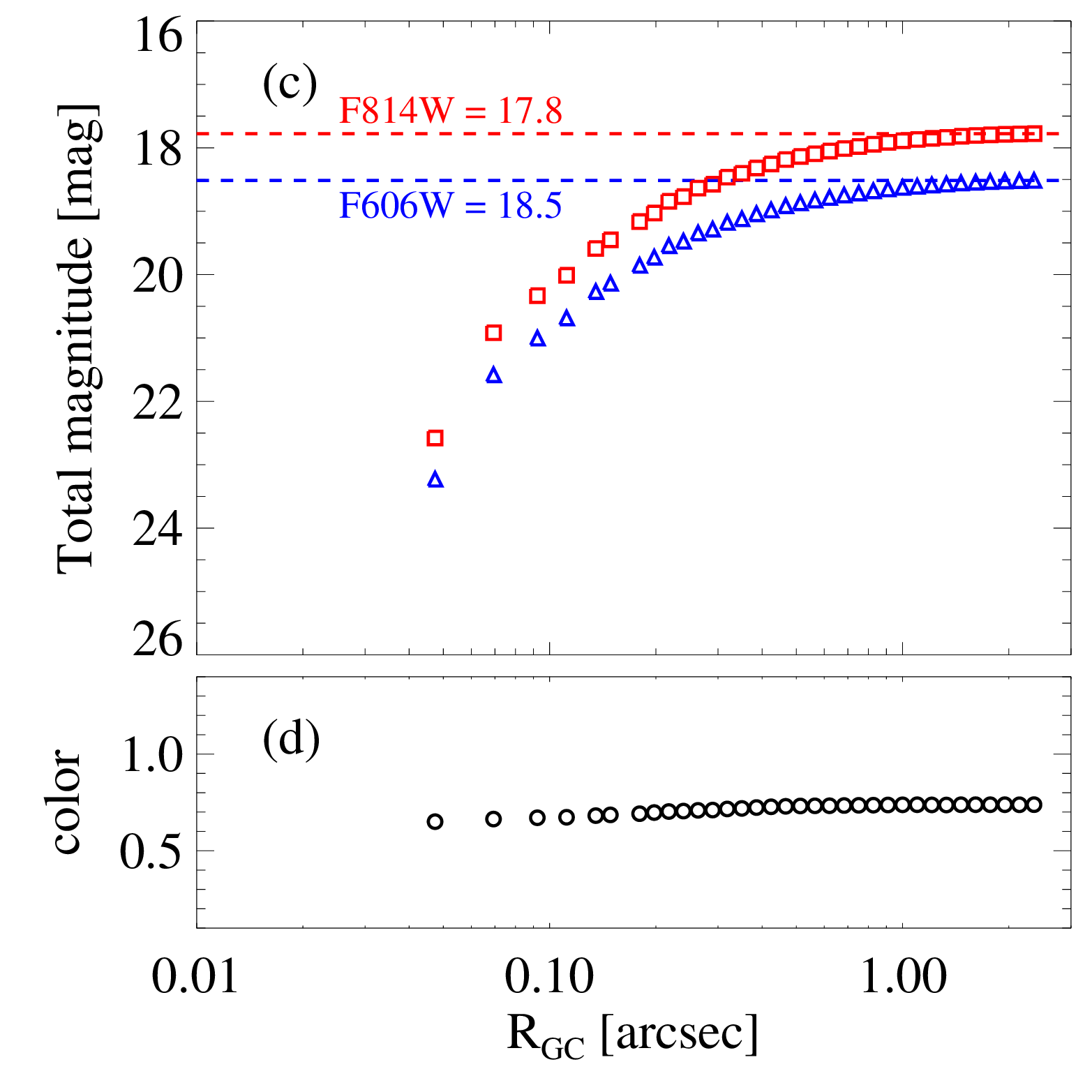} 
\caption{
(a) Surface brightness profiles of M64-GC1 obtained using \texttt{IRAF/ELLIPSE}
(blue triangles for $F606W$, and red boxes for $F814W$). 
Dotted lines represent King profile fitting. 
(b) $(F606W-F814W)$ color profile of M64-GC1. 
Its color is almost constant with $(F606W-F814W)\approx$ 0.7 to 0.8 at $R_{GC}<1\farcs5$.
(c) Integrated magnitude profiles of M64-GC1. 
Total magnitudes are $F606W=18.5$ mag and $F814W=17.8$ mag (marked by dashed lines). 
(d) Integrated $(F606W-F814W)$ color profile of M64-GC1.
}
\label{fig_fig6}
\end{figure*}

In {\color{blue}\bf Figure \ref{fig_fig5}(a)} we display the CMD of the resolved stars in M64-GC1. 
We plotted the stars at $1\farcs2<R_{GC}<2\farcs7$, 
most of which are considered to be cluster members, by magenta circles.
Then we plotted the stars in the outer background region at $4\farcs0<R_{GC}<7\farcs0$, 
which must be background sources outside the cluster, by pale blue circles.
The area of the outer background region is about six times larger than that of the central cluster region.
For comparison, we overlayed \texttt{PARSEC} isochrones \citep{bre12} 
for ages of 12 Gyr and [Fe/H] = --2.2 to 0.0 with a step of 0.2 dex (green lines).
These isochrones are shifted according to the distance ($(m-M)_0=28.2$) and foreground reddening ($A_{F606W}=0.102$ and $A_{F814W}=0.063$) of M64 \citep{sch11}.

In the CMD, most stars in the cluster region are located along a vertical branch at $(F606W-F814W)\approx 1.0$, 
which is consistent with RGB isochrones with low metallicity, [Fe/H] $\approx -1.5$. 
In contrast, the stars in the background region are seen along curved branches with much redder colors.
Therefore, the stars in the cluster region must be mostly the members of the new metal-poor globular cluster. 
We estimate the mean metallicity of the cluster, from the comparison of these RGB stars with 12 Gyr isochrones, obtaining [Fe/H] $=-1.5\pm0.2$.
This shows that M64-GC1 is indeed an old metal-poor globular cluster belonging to the halo population.

For comparison, we display the CMD of the resolved stars in Coma P, a new dwarf companion of M64, in {\color{blue}\bf Figure \ref{fig_fig5}(b)}.
\citet{bru19} presented $F606W/F814W$ photometry of the resolve stars in Coma P.
To obtain the CMD of the resolved stars in Coma P, we applied a similar reduction and photometry process for the {\it HST}/ACS $F606W/F814W$ images of Coma P (PI: Salzer, PID: 14108) as used for M64 images.
The CMD of Coma P we obtained is consistent with the result of \citet{bru19}.
The same isochrones in {\color{blue}\bf Figure \ref{fig_fig5}(a)} are overlayed after a shift according to the distance ($(m-M)_0=28.7$) and foreground reddening ($A_{F606W}=0.077$ and $A_{F814W}=0.048$) of Coma P given by \citet{bru19}.
Unlike the RGB stars in M64-GC1, the RGB stars of Coma P show a much broader color (metallicity) distribution
with a slightly higher mean metallicity, [Fe/H] $\approx -0.9$.
Indeed, M64-GC1 has a lower metallicity than Coma P.
Moreover, there are a dozen young stars located at $(F606W-F814W)\approx 0$ in the CMD of Coma P,
which are not seen in the CMD of M64-GC1.

We estimate the total integrated magnitude and effective radius of M64-GC1 
from surface photometry using \texttt{IRAF/ELLIPSE} task on the ACS images \citep{jer87}.
{\color{blue}\bf Figures \ref{fig_fig6}(a) and (b)} display the surface brightness and color profiles of M64-GC1.
We fit the surface brightness profiles of the cluster with a King profile
\citep{kin62}, as shown by dotted lines in {\color{blue}\bf Figure \ref{fig_fig6}(a)}.
We derive 
core radius $r_c=0\farcs13\pm0\farcs01$ ($2.7\pm0.2$ pc) and tidal radius $r_t=2\farcs59\pm0\farcs02$ ($54.4\pm0.4$ pc) in $F814W$ images, and
$r_c=0\farcs12\pm0\farcs01$ ($2.5\pm0.2$ pc) and $r_t=2\farcs64\pm0\farcs02$ ($55.4\pm0.4$ pc) in $F606W$ images.
These values are obtained after correcting the PSF effect.
The color profile is almost constant with $(F606W-F814W)\approx$ 0.7 to 0.8 at $R_{GC}<1\farcs5$.

The integrated magnitude and color profiles of the cluster are shown in {\color{blue}\bf Figures \ref{fig_fig6}(c) and (d)}.
From this we derive the total magnitude of the cluster,
$F606W(GC1)=18.5$ mag, and $F814W(GC1)=17.8$ mag. 
We also fit the S\'ersic profile to the M64-GC1 image using \texttt{GALFIT} \citep{pen10}.
From this we derive the total magnitude of the cluster,
$F606W(GC1)=18.53\pm0.01$ mag, and $F814W(GC1)=17.80\pm0.01$ mag 
(corresponding to the Johnson Cousins system magnitudes, $V_0=18.64\pm0.03$ mag and $I_0=17.72\pm0.02$ mag),
and the effective radius of R$_{\rm eff}=0\farcs273\pm0\farcs001$ ($5.73\pm0.02$ pc).
These results agree very well with those from the \texttt{IRAF/ELLIPSE} task.
Adopting the distance to M64, we obtain absolute integrated magnitudes of the cluster,
$M_{F606W}(GC1)=-9.75\pm0.09$ mag, and $M_{F814W}(GC1)=-10.44\pm0.09$ mag 
(corresponding to the Johnson Cousins system magnitudes, $M_V=-9.54\pm0.09$ mag and $M_I=-10.46\pm0.09$ mag). 
In {\color{blue}\bf Table \ref{tab_gc1}}
we list the basic information of M64-GC1 derived in this study.

\begin{deluxetable}{lcc}[hb!]
\tablecaption{Basic Parameters of M64-GC1 \label{tab_gc1} }
\tablewidth{0pt}
\tablehead{
\colhead{Parameter} & \colhead{Value} 
}
\startdata
R.A.(2000) & 12$^h$ 57$^m$ 04$^s$.58 \\
Decl.(2000) & +21$\arcdeg$ 38$\arcmin$ 23$\farcs$7 \\
$F606W$ & $18.53\pm0.01$ \\
$F814W$ & $17.80\pm0.01$ \\
$F606W-F814W$ & $0.73\pm0.01$ \\
$V_0$ & $18.64\pm0.03$ \\
$(V-I)_0$ & $0.92\pm0.13$ \\
$M_V$ & $-9.54\pm0.09$ \\
$R_{\rm eff}$ & $0\farcs273\pm0\farcs001$ \\
 & ($5.73\pm0.02$ pc) \\
Mean RGB metallicity & [Fe/H]$=-1.5\pm0.2$ \\
\enddata
\end{deluxetable}


\subsection{CMDs of the Resolved Field Stars}

\begin{figure} 
\centering
\includegraphics[scale=0.48]{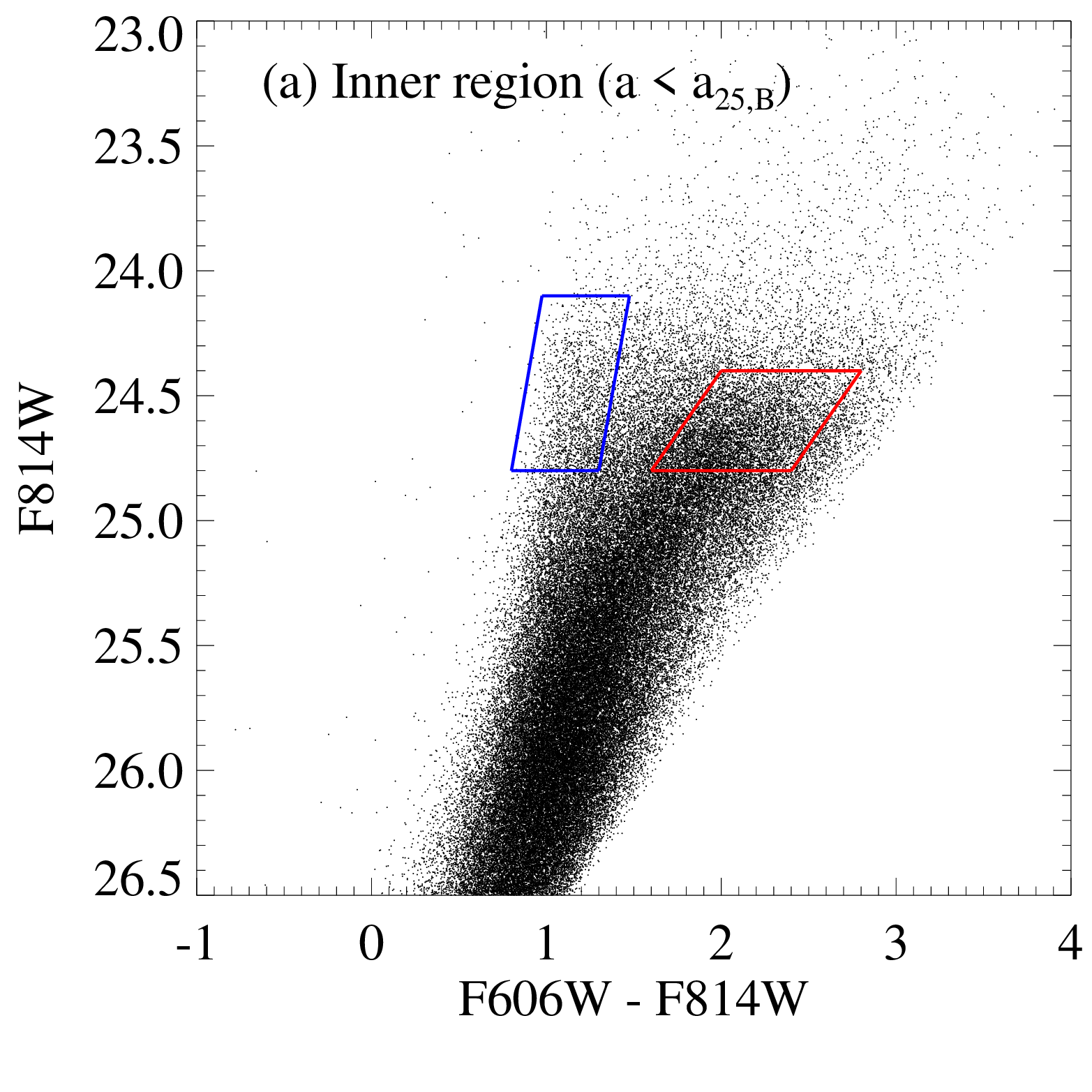}
\includegraphics[scale=0.48]{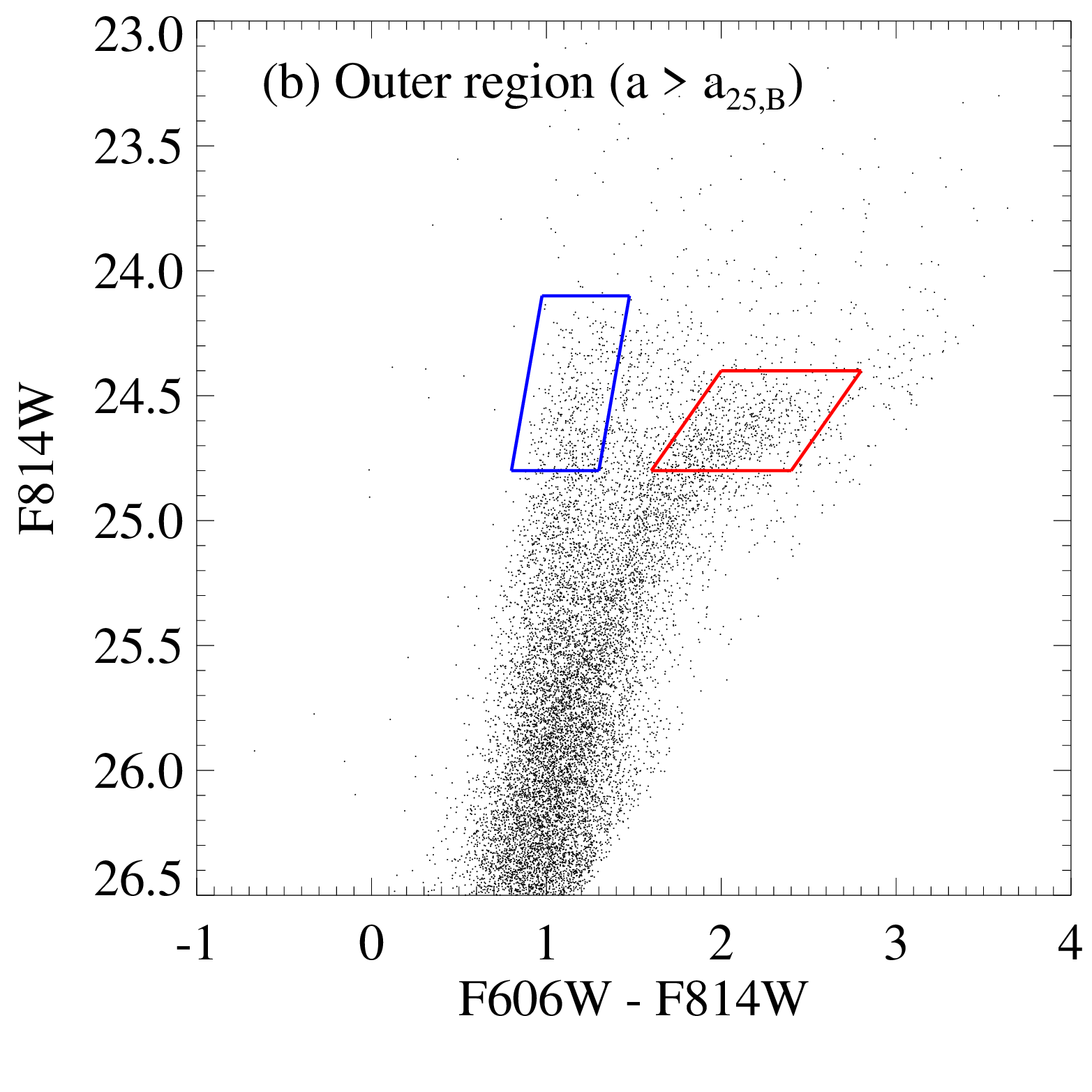}
\caption{
CMDs of the resolved stars (a) in the inner region at $a<a_{25,B}=5\farcm0$ and (b) the outer region at $a\ge a_{25,B}$.
Blue and red boxes denote the boundaries for selecting 
bright metal-poor RGB stars and bright metal-rich RGB stars, respectively.
}
\label{fig_fig7}
\end{figure}

For the following analysis we divided the entire ACS field into two regions: 
an inner region at $a<a_{25,B}= 5\farcm0$ (6.3 kpc) and an outer region at $a\geq a_{25,B}$.
Here, 
$a_{25,B}$ is the projected galactocentric distance at which the $B$-band surface brightness is 25 mag arcsec$^{-2}$.
$a_{25,B}$ is shown by a large ellipse in {\color{blue}\bf Figure \ref{fig_fig1}}.
In {\color{blue}\bf Figure \ref{fig_fig7}} we display the CMDs of the resolved stars in the inner and outer regions. 
It is found that the resolved stars are mostly RGB stars with a large range of color.

The most notable feature in the CMDs is that there are two distinguishable RGBs in the bright magnitudes with $F814W \lesssim 25.0$:
one is a vertical blue branch at $(F606W-F814W) \approx 1.0$, 
and the other is a curved red branch with a wide range of color reaching $(F606W-F814W) \lesssim 2.5$.
The vertical RGB represents a metal-poor RGB, 
and the curved RGB denotes a metal-rich RGB.
The presence of these two branches is more clearly seen in the outer region than in the inner region.
The inner region is dominated by the metal-rich RGB stars, 
while the contribution from the metal-rich RGB stars and metal-poor RGB stars are comparable in the outer region.

The metal-poor RGB in the CMDs is very similar to the RGB of the new globular cluster M64-GC1, 
showing that the metal-poor RGB stars belong to the halo population.
In contrast, the metal-rich RGB is roughly matched with the isochrone for [Fe/H] $\approx -0.4$, 
which is much larger than the mean metallicity of the metal-poor RGB ([Fe/H] $\approx -1.5$). 
Thus, the metal-rich RGB represents the disk population. 

The red color 
of the outer disk presented in \citet{wat16} is mainly due to these metal-rich RGB stars. 
However, taking advantage of the resolved stellar photometry, 
we find that the red outer disk is not made of a single stellar population with high metallicity, 
but it is composed of two stellar populations: 
one with high metallicity and another with much lower metallicity.

\subsection{TRGB Distance Estimation}

\begin{figure} 
\centering
\includegraphics[scale=0.48]{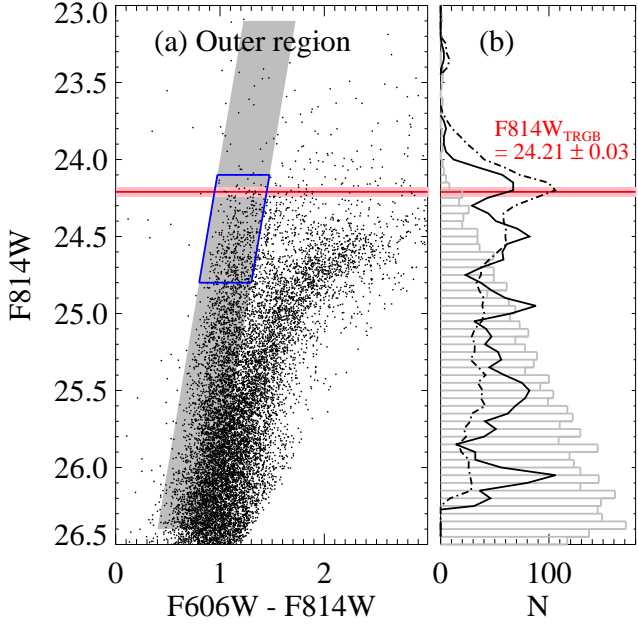}
\caption{
(a) CMD of the resolved stars in the outer region at $a\ge a_{25,B}=5\farcm0$.
The shaded grey region represents the stars used for estimating the RGB tip.
(b) $F814W$ band luminosity function of the metal-poor RGB stars in the outer region. 
The edge detection response curve is also plotted 
(a solid line for the Sobel filter and dashdotted line for the weighted edge detection filter suggested by \citet{mag08}).
Red lines mark the TRGB magnitude, $F814W_{TRGB} = 24.21\pm0.03$. 
}
\label{fig_fig8}
\end{figure}

We estimate a distance to M64 using the TRGB method \citep{lee93} 
to a sample of metal-poor RGB stars in the outer region of the ACS field.
Crowding is much less and the contribution of halo stars is larger in the outer region than in the inner region.
{\color{blue}\bf Figure \ref{fig_fig8}(a)} is the CMD of the resolved stars in the outer region of M64 
which is already shown in {\color{blue}\bf Figure \ref{fig_fig7}(b)}.
The shaded grey region represents the metal-poor RGB stars in the outer region 
which were used to estimate the TRGB magnitude.

{\color{blue}\bf Figure \ref{fig_fig8}(b)} shows the luminosity function of the stars within the shaded grey region in the CMD. 
Applying the Sobel filter method described in \citet{jan17} to the luminosity function of the stars,
we derive a TRGB magnitude, $F814W_{TRGB}=24.21\pm0.03$ mag.
This TRGB magnitude does not change much with the choice of Sobel filters. 
For example, the edge detection algorithm used in \citet{fre19}, employing the GLOESS smoothing, 
gives the TRGB magnitude of $F814W_{TRGB}=24.22\pm0.02(\rm ran)\pm0.01(\rm sys)$ mag 
with the optimal smoothing scale of $\sigma_s=0.08$ mag. 
The weighted edge detection filter suggested by \citet{mag08} measures $F814W_{TRGB}=24.21\pm0.04$ mag, 
which is also consistent within uncertainties.
Adopting the TRGB calibration in \citet{jan17} (for metal-poor RGB, $M_{F814W}(TRGB) = -4.03 \pm 0.07$), 
the foreground reddening in \citet{sch11} ($A_{F814W}=0.063$), 
and the aperture correction error of 0.03 mag,
we obtain a distance modulus,
$(m-M)_0=28.18\pm0.03{\rm (ran)}\pm0.08{\rm (sys)}$ (corresponding to a distance of $4.33\pm0.18$ Mpc).

\subsection{Radial Distribution of the Metal-poor and Metal-rich RGB Stars}\label{rdp}

In the CMDs of {\color{blue}\bf Figure \ref{fig_fig7}}, 
the metal-poor RGB and the metal-rich RGB are clearly separated in the bright magnitudes.
They are merged into one as the stars get fainter so it is difficult to separate them in the faint magnitudes.
For the following analysis we selected bright metal-poor RGB stars and metal-rich RGB stars with $F814W < 24.8$ mag using the boundaries marked by blue and red boxes in 
{\color{blue}\bf Figure \ref{fig_fig7}}.

\begin{figure} 
\centering
\includegraphics[scale=0.48]{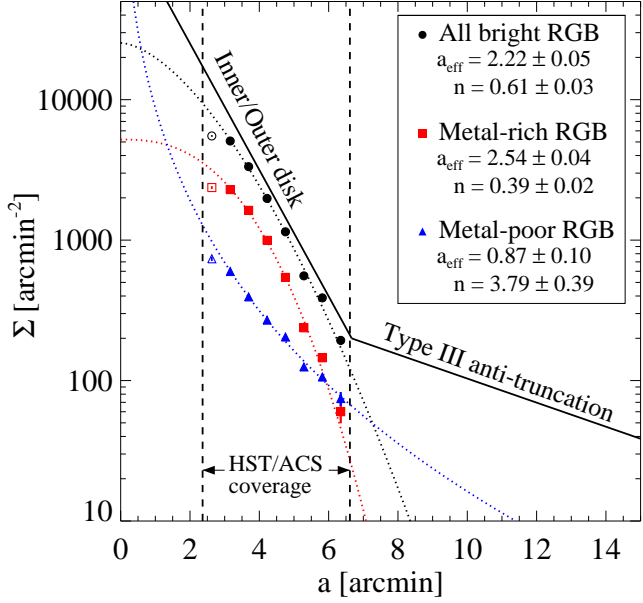} 
\caption{
Radial number density profiles of the selected metal-poor RGB stars (blue triangles), metal-rich RGB stars (red boxes), and all bright RGB stars (black circles) with $24.1<F814W<24.8$ mag.
The surface brightness profile of galaxy light given by \citet{wat16} is schematically shown: black solid line for the inner/outer disk and the Type III anti-truncation component.
Vertical dashed lines mark the {\it HST}/ACS coverage and dotted lines denote S\`ersic fitting lines.
The fitting results are also shown.
}
\label{fig_fig9}
\end{figure}

We derive the radial number density profiles of the selected metal-poor RGB stars and metal-rich RGB stars 
as well as all bright RGB stars with $24.1<F814W<24.8$ mag, 
and show them in {\color{blue}\bf Figure \ref{fig_fig9}}.
This result was corrected by the artificial star tests (see {\color{blue}\bf Figure \ref{fig_fig2}}).
For comparison, we also plot the schematic surface brightness profiles of galaxy light 
given by \citet{wat16}, showing the inner/outer disk and the Type III anti-truncation component.
We converted the schematic surface brightness profiles to be consistent with the unit of the number density profiles, $\mu_B$ divided by 2.5, and shifted them vertically to match roughly the number density profile at $a \approx 400''$.

Several features are noted in {\color{blue}\bf Figure \ref{fig_fig9}}.
First, the radial number density profile of the metal-rich RGB stars 
is much steeper than that of the metal-poor RGB stars in the same range.
The flattening or decrease of the  radial number density profiles in the innermost bins at $a<3\arcmin$ are mainly due to incompleteness of our photometry
although we corrected the incompleteness with the result of artificial star tests.
Second, the slope of the radial number density profile of the metal-rich RGB stars is consistent with that of the surface brightness profile for the outer disk.
Third, the slope of the radial number density profiles of all bright RGB stars is similar to that of the surface brightness profile for the outer disk.
Fourth, the radial number density profile of the metal-poor RGB stars is slightly steeper than, 
but roughly consistent with that of the surface brightness profile for the Type III component given by \citet{wat16}.

We fit the radial number density profiles for $3'\lesssim a \lesssim 6'$ with a S\'ersic function \citep{ser63},
obtaining
$n=0.61\pm0.03$ ($a_{\rm eff}=2\farcm22\pm0.05$) for all bright stars,
$n=0.39\pm0.02$ ($a_{\rm eff}=2\farcm54\pm0.04$) for the metal-rich RGB stars, and
$n=3.79\pm0.39$ ($a_{\rm eff}=0\farcm87\pm0.10$) for the metal-poor RGB stars.
Thus, the radial number density profile of the metal-rich RGB stars follows an exponential disk law, 
while that of the metal-poor RGB stars is described well by a de Vaucouleurs $r^{1/4}$ law.

\section{Discussion}

\subsection{M64-GC1 in Comparison with Other Globular Clusters}

\begin{figure} 
\centering
\includegraphics[scale=0.48]{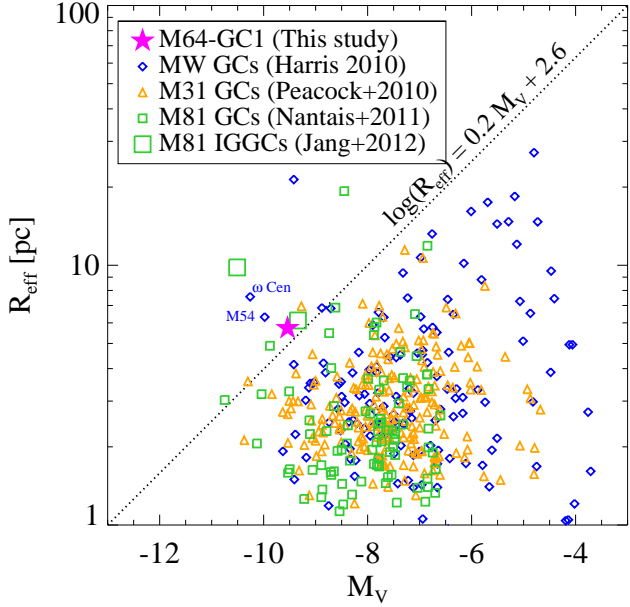} 
\caption{
Effective radius ($R_{\rm eff}$ [pc]) versus $M_V$ for M64-GC1 (magenta starlet) in comparison with globular clusters in the Milky Way Galaxy (blue diamonds) \citep{har96,har10}, M31 (orange triangles) \citep{pea10}, M81 (green boxes) \citep{nan11}, and two intragroup globular clusters (IGGCs) in the M81 group (large green boxes) \citep{jan12}.
A boundary separating normal and extended globular clusters suggested by \citet{van04} is marked by a dotted line.
}
\label{fig_fig10}
\end{figure}

It is easy to spot M64-GC1 in the ACS images because it is relatively bright and its outer region is partially resolved into individual stars.   
However, it is the only globular cluster we could find in the ACS field.
This implies that the total number of globular clusters in M64 is small and/or 
M64-GC1 may be unusually brighter than the other globular clusters in this galaxy.

In {\color{blue}\bf Figure \ref{fig_fig10}} we compare the effective radius and total magnitude of M64-GC1 with those of globular clusters in other galaxies in the literature (the Milky Way Galaxy \citep{har96,har10}, M31 \citep{pea10}, and M81 \citep{nan11}) and two intragroup globular clusters (IGGCs) in the M81 group \citep{jan12}. 
We also plotted a boundary separating normal and extended globular clusters suggested by \citet{van04}: log($R_{\rm eff} ) = 0.2 M_V + 2.6$. 
M64-GC1, as well as M54 (NGC 6715), $\omega$ Centauri (NGC 5139), and IGGCs, is above the boundary line. 
This shows that M64-GC1 belongs to the brightest and largest globular clusters.
M54 and $\omega$ Centauri are often considered to be core remnants of disrupted dwarf galaxies 
\citep{zin88,fre93,sar95}. 
Therefore, M64-GC1 is also a possible core remnant of a tidally disrupted dwarf satellite around M64.


\subsection{Comparison of Distance Estimation}

\begin{deluxetable*}{lccll} 
\tabletypesize{\footnotesize}
\tablecaption{Distance Estimates for M64 \label{tab_distance} }
\tablewidth{0pt}
\tablehead{
\colhead{Method} & \colhead{D[Mpc]} & \colhead{$(m-M)_0$} & \colhead{Reference} & \colhead{Remarks} 
}
\startdata
SBF & 7.48 & $29.37\pm0.20$ & \citet{ton01}  & $I$-band \\
SBF & 5.24 & $28.60\pm0.08$ & \citet{tul13}  & \\
\hline
TRGB & 5.27 & $28.61$ & \citet{mou08} & $F814W_{TRGB,0}=24.64$\\
TRGB & 4.66 & $28.34\pm0.01$ & \citet{jac09} & $F814W_{TRGB,0}=24.49$ \\ 
 &  &  &  & Adopted in \citet{wat16} \\
TRGB & 5.24 & $28.60\pm0.16$ & \citet{tul13} & \\
TRGB & 5.30 & $28.62\pm0.15$ & \citet{tul16} & Adopted in \citet{bru19} \\
TRGB & 4.40 & $28.22\pm0.02$ & EDD (2019)& $F814W_{TRGB,0}=24.26\pm0.02$ \\ 
TRGB & 4.33 & $28.18\pm0.09$ & This study & $F814W_{TRGB,0}=24.15\pm0.03$ \\ 
\enddata
\end{deluxetable*}

The previous TRGB distance estimates for M64 show a significant variation from $(m-M)_0=28.2$ to 28.6, 
as summarized in {\color{blue}\bf Table \ref{tab_distance}} \citep[][and the Extragalactic Distance Database (EDD)\footnote{\url{http://edd.ifa.hawaii.edu/}}]{mou08,jac09,tul13,tul16}. 
The previous surface brightness fluctuation (SBF) distance values show an even larger difference: $(m-M)_0=29.37\pm0.20$ in \citet{ton01} and $(m-M)_0=28.60\pm0.08$ in \citet{tul13}. 
Note that \citet{wat16} adopted a distance 4.7 Mpc given in \citet{jac09}, while \citet{bru19} adopted a distance 5.3 Mpc given in \citet{tul16}.

We obtain a distance to M64 applying the TRGB method to the photometry of metal-poor RGB stars in the outer region, $(m-M)_0=28.18\pm0.09$ ($4.33\pm0.18$ Mpc).
This value is 0.4 mag smaller than those in \citet{mou08}, \citet{tul13}, and \citet{tul16}, but it is very similar to the one in EDD, $(m-M)_0=28.22\pm0.02$.

Note that the $F814W$ TRGB magnitude in this study, $F814W_{TRGB,0}=24.15\pm0.03$, is 0.5 mag brighter than the value in \citet{mou08}, 24.64 mag. 
We used only the metal-poor RGB stars in the outer region, while \citet{mou08}'s (as well as \citet{tul13}'s and \citet{tul16}'s) estimation might have been based mainly on all RGB stars.
If we determine the TRGB magnitude using all RGB stars, 
we obtain $F814W_{TRGB,0}\approx24.5$ mag, which is similar to those obtained by \citet{mou08} and \citet{jac09} (probably as well as \citet{tul13} and \citet{tul16}).

EDD provided a value of $F814W_{TRGB,0}=24.26\pm0.02$, which is 0.1 mag fainter than ours and presented a similar distance value.
If we determine the TRGB magnitude using metal-poor RGB stars in the entire region,
we obtain $F814W_{TRGB,0}\approx24.25$, which is similar to the value provided by EDD.
However, they used all RGB stars in the entire region without any color selection (private communication with Dmitry Makarov). Then it is difficult to explain the 0.1 mag difference.

\subsection{The Origin of the Type III Component in M64}

\citet{erw05,erw08} and \citet{poh06} introduced the classification of Type III breaks into two types
according to the shape of outer isophotes: 
IIId for disk components and IIIs for spheroidal components. 
In the case of IIIs breaks, the isophotes in the outer region become progressively rounder than those in the inner region. 
They described the spheroidal component in IIIs breaks is either the outer extent of the bulge or a separate stellar halo.

From the surface photometry of M64, \citet{gut11} classified it as IIIs, 
noting that there is a disk break at $r\approx370''$ 
and that the ellipticity of the isophotes decreases abruptly beyond $r\approx340''$. 
\citet{wat16} also found a disk break at $r\approx400''$ and inferred 
that the Type III anti-truncation component in M64 may be due to the spheroidal halo, 
noting that the isophotes of the outer region of M64 are round
and that it is difficult to build outer disks with current models. 


We find that the resolved RGB stars in the ACS field are composed of two main stellar populations: 
metal-poor RGB stars and metal-rich RGB stars.
The radial number density profile of the metal-rich RGB stars is described by an exponential disk law, 
while that of the metal-poor RGB stars is fit well by a S\'ersic law with $n\approx4$.
In the CMDs, the metal-poor RGB is very similar to the RGB of the new halo globular cluster (M64-GC1).

These results show that the metal-poor RGB represents a low metallicity ([Fe/H] $\approx -1.5$) halo population, while the metal-rich RGB denotes a disk population with much higher metallicity ([Fe/H] $\approx -0.4$).
The disk is dominated by the metal-rich RGB stars, 
and no clear sign of young stars is found in the ACS field.
Thus, the color of the outer disk is dominated by the contribution of the metal-rich RGB stars.
This is consistent with the result that the outer disk is red and featureless found in \citet{wat16}.
The outer region is occupied mainly by the metal-poor RGB stars. 
Thus, the origin of the Type III anti-truncation component in the outer region of M64 is a halo, 
neither a disk population nor a bulge population.


\subsection{The Origin of the Halo in M64}

M64 is an 
isolated galaxy located in a low density environment \citep[see][Figure 10]{bru19}.
In such an environment, there might have been few satellites which merged and enlarged the halo of M64. 
This implies that the size and mass of the halo in M64 might have not been changed much 
since its formation, or the accretion of dwarf galaxies into M64 was finished long ago. 
Thus, the halo in M64 may be pristine, 
as dwarf galaxies or globular clusters which have much lower mass than M64. 
This is consistent with our finding that the metallicity of the metal-poor RGB stars in the ACS field 
is as low as that of the new halo globular cluster (M64-GC1).
The metallicity of the metal-poor RGB stars is much lower than that of the metal-rich RGB stars 
so the origin of the metal-poor RGB stars cannot be a disk or a bulge.

\subsection{Relation between M64 and Coma P}

Coma P (AGC 229385) is an HI-dominated blue dwarf galaxy 
($M(HI)=3.48\times10^7M_\odot$, $M_*=4.3\times10^5M_\odot$)
located close ($5.86^\circ$) to M64 in the sky \citep{bru19}.
\citet{bru19} measured a TRGB magnitude of Coma P from {\it HST}/ACS $F606W/F814W$ images 
to be $F814W_0= 24.64\pm0.09$ (from maximum likelihood method) or 24.60 (from Sobel filters), 
and presented a distance to Coma P,  $d=5.5^{+0.28}_{-0.53}$ Mpc 
($(m-M)_0=28.80^{+0.11}_{-0.21}$). 
This value is much smaller than the previous distance estimates for Coma P, 
11 Mpc based on the TRGB method \citep{ana18} and 25 Mpc based on the flow model \citep{jan15}.
\citet{bru19} discussed the comparison of their distance estimate with those in \citet{jan15} and \citet{ana18}, 
and concluded that their estimate is more reliable than the others. 

\citet{bru19} discussed the relation between Coma P and M64, 
adopting a distance to M64 from Cosmicflows-3 in \citet{tul16}, $(m-M)_0=28.62\pm0.15$ ($d=5.3\pm0.4$ Mpc). 
The spatial distance between the two galaxies is as small as 590 kpc for their adopted distances. 
However, the relative line-of-sight velocity between Coma P ($v_{helio}=1348$ \kms) and M64 ($v_{helio}=410$ \kms) is as large as 938 \kms.
Noting this large relative velocity, \citet{bru19} suggested another scenario for the presence of counter-rotating gas in M64 in relation with Coma P:
a gas-rich progenitor of Coma P had a high speed ($v\approx$ 1000 \kms~ $\approx$ 1 Mpc Gyr$^{-1}$) fly-by interaction with M64 approximately 1 Gyr ago, 
and some of the gas stripped from Coma P resulted in the counter-rotating gas ring in the outer disk of M64.

If the TRGB distance to M64 derived in this study is adopted, 
then the relative line-of-sight distance between Coma P and M64 increases from 200 kpc to 1.2 Mpc 
(corresponding to the spatial distance from 590 kpc to 1.3 Mpc).
It would have taken about 1 Gyr if Coma P had an encounter with M64 and is receding with $v\approx 1000$ \kms~ until today.
Therefore, the encounter epoch is around 1 Gyr ago, which is in agreement with the scenario suggested by \citet{bru19}.

\subsection{Evolution History of M64}

To explain the counter-rotating gas in the outer disk of M64, 
it has been suggested that the progenitor spiral galaxy of M64 underwent a merger with a counter-rotating gas-rich galaxy
\citep{bra92,bra94,rub94,wal94,rix95}. 
Based on the result that only the outer gas disk is counter-rotating 
while the outer stellar disk is co-rotating with the inner gas disk and stellar disk, 
they prefer mergers with a low-mass gaseous system.

Alternatively, they also suggested a continuous accretion of gas with an opposite sense of angular momentum. 
Several studies also showed that a majority of isolated S0 galaxies host extended counter rotating gas disks 
originated from external gas \citep{dav11,kat14,kat15}. 
Especially, \citet{kat14} mentioned that possible sources of external gas accretion 
are systems of dwarf gas-rich satellites or cosmological cold-gas filaments \citep{ker05,dek06}.

Recently, \citet{wat16} supported a merger scenario again to explain the presence of young stars in the inner disk, 
a red and featureless outer disk, and counter-rotating gas in the outer disk of M64. 
The recent star formation is limited only in the inner disk, while the outer disk is quietly evolving. 
Thus, \citet{wat16} pointed out that the merger played two opposite roles: 
it induced star formation in the inner region, while it removed gas and quenched star formation in the outer disk. 
They also suggested that M64 may be at the ending phase of a spiral galaxy being transformed to an S0 galaxy by merger-induced quenching. 
Indeed, the color of M64 is as red as S0a galaxies \citep{rob94}
and Type III anti-truncations are more frequently seen in S0 galaxies \citep{bor14}. 

More recently, \citet{bru19} suggested another scenario that M64 had experienced a fly-by interaction 
with a gas-rich galaxy rather than a merger. 
They suggested that a gas-rich progenitor of Coma P had a high speed fly-by interaction with M64 approximately 1 Gyr ago, 
and some of the gas stripped from Coma P resulted in the counter-rotating gas ring in the outer disk of M64. 
They also pointed out that young stars in Coma P seen in the CMD might have formed during the interaction of Coma P with M64.

From all the previous results, we support the scenario suggested by \citet{bru19}. 
The presence of old RGB stars in both M64 (in this study) and Coma P \citep{bru19} implies 
that both galaxies were formed more than 10 Gyrs ago. 
The presence of young stars in Coma P and the central region of M64 
strongly indicates a recent interaction between the two galaxies. 
The counter-rotating gas ring with clumpy structure in M64 must have an external origin: 
either via accretion or fly-by interaction of dwarf gas-rich satellites, or cosmological cold-gas filaments. 
Coma P is a strong candidate for providing the counter-rotating gas to M64 through fly-by interaction about 1 Gyr ago. 


\section{Summary and Conclusion}

We presented deep $F606W/F814W$ photometry of resolved stars in a field located in the outer disk of M64.
We used it to explore the nature of stellar populations in the outer disk and find a clue to reveal the origin of the Type III component.
Primary results are summarized as follows.

\begin{itemize}

\item We discovered a new globular cluster, M64-GC1. The CMD of this cluster is well matched with 12 Gyr isochrones for low metallicity [Fe/H]$=-1.5\pm0.2$, showing that it is a halo population.

\item CMDs of the resolved stars in the ACS field show two distinguishable stellar populations: metal-poor RGB stars and metal-rich RGB stars.
In the CMDs, the metal-poor RGB is very similar to the RGB of the new halo globular cluster (M64-GC1), showing that the metal-poor RGB stars belong to the halo.

\item The radial number density profile of the metal-rich RGB is described by an exponential disk law, while that of the metal-poor RGB is fit well be a de Vaucouleurs's law.
This shows that the metal-rich RGB represents the disk population with high metallicity, while the metal-poor RGB denotes the halo population with low metallicity.

\item We derive a TRGB distance to M64, $d=4.33 \pm 0.18$ Mpc ($(m-M)_0 = 28.18 \pm 0.03(\rm ran) \pm 0.08(\rm sys)$.

\end{itemize}

From these results, we conclude that the origin of the Type III break in M64 is a halo 
rather than a disk or a bulge. 

\acknowledgements

J.K. was supported by the Global Ph.D. Fellowship Program (NRF-2016H1A2A1907015) of the National Research Foundation (NRF).
This project was supported by the National Research Foundation grant funded by the Korean Government (NRF-2019R1A2C2084019).
We thank Brian S. Cho for his help in improving the English in the draft. 
We also thank the anonymous referee for useful suggestions to improve the original manuscript.

\vspace{5mm}
\facility{HST(ACS)}

\software{\texttt{DrizzlePac} \citep{gon12},
        \texttt{DOLPHOT} \citep{dol00},
        \texttt{IRAF/ELLIPSE} \citep{jer87},
        \texttt{GALFIT} \citep{pen10}
          }

.

\end{document}